\documentstyle[prl,aps,epsf]{revtex}

\begin{document}
\title{Geometric chaoticity leads to ordered spectra for randomly interacting fermions}
\author{D. Mulhall, A. Volya and V. Zelevinsky }
\address{Department of Physics and Astronomy and National Superconducting Cyclotron Laboratory, Michigan State University,\\ East Lansing, Michigan 48824-1321}

\date{\today}
%MSUCL-1151  preprint number.
\maketitle

\begin{abstract}
   A rotationally invariant random interaction ensemble was realized in a single-$j$ fermion model. The dominance of ground states with zero and maximum spin  was confirmed and explained with a statistical approach based on the random coupling of individual angular momenta. The interpretation is supported by the structure of the ground state wave functions.
\end{abstract}

The interplay of regular and chaotic features in many-body quantum dynamics is
currently extensively studied both for simple models and for realistic
applications to atomic \cite{grib}, nuclear \cite{big,ann,guhr}, and condensed 
matter physics \cite{alt?}, as well as for understanding properties of the QCD 
vacuum \cite{verbaar}. Typical finite ``shell-model" systems such as complex atoms  and nuclei are described by the mean field and corresponding residual
interaction. The density of the mean field configurations grows exponentially
for combinatorial reasons, so that the interaction becomes effectively 
strong at sufficiently high excitation energy leading to generic chaotic features both in spectral statistics, which 
rapidly move to the limit of random matrix theory \cite{guhr,brody}, and in 
properties of wave functions \cite{grib,big}. Studies of finite many-body systems have to account for the existence of
constants of motion such as total angular momentum, isospin and parity. If these conservation laws are exact, one usually deals with the states of each class separately.  However, little attention was paid to the problem of
correlations between classes of states which are described by the same 
Hamiltonian but belong to different values of exact integrals of motion.

An obvious and practically important example is  angular momentum 
conservation in a finite Fermi-system. The
prediagonalization procedure of projecting the correct value $J$ of nuclear 
spin out of the $m$-scheme Slater determinants induces by itself a strong
mixing of the states within a shell model configuration \cite{big}. The projected states
of various spins acquire a nearly uniform degree of complexity and energy
dispersion. For a sufficiently large dimension, the majority of
states correspond to a complicated quasi-random coupling of individual 
spins. This ``geometric chaoticity" was used long ago \cite{erik} in
evaluating the level density for a given $J$. It also plays an
important role in the response to external fields, large
amplitude collective motion, dissipation and so on \cite{ann}. 
The similarity of different
$J$-classes with respect to mixing was demonstrated \cite{baps,m-b}
in the nuclear shell model
by the studies of complexity, occupation numbers,
strength functions and pairing properties. This
raises also a question of existence of compound rotational bands \cite{doss}
which would connect  complicated states having different $J$ but almost the
same mixing.

A new angle of looking at the problem was introduced by refs. \cite{b1,b2}
where the spectrum of a random but rotationally invariant Hamiltonian was
obtained for a shell-model Fermi system. In spite of the random character of
the two-body interaction, the fraction $f_{0}$
of the ensemble realizations with a ground state spin  $J_{0}=0$ 
was much higher than the total statistical fraction $f^{s}_{0}$
of $J=0$ states in shell-model space. This result was confirmed in refs. \cite{baps2,stu}, as well as for the interacting boson model
\cite{bos1}. A new feature discovered in \cite{baps2,bos1} was an excess of the probability $f_{J_{max}}$ for the ground state to have
the maximum possible
spin $J_{max}$. The emergence of regular features as an output of a random
interaction seems to contradict the notion of geometrical chaoticity.
Below we show that, vice versa, the geometric chaoticity provides a base
for explaining the main features of the pattern.

First we give a couple of trivial examples which point out the possible source
of the effects, namely an analog of the Hund rule in atomic
physics. Consider a system of $N$ pairwise interacting spins with the
Hamiltonian
\begin{equation}
H=A\sum_{a\neq b}{\bf s}_a\cdot{\bf s}_{b}=A[{\bf S}^{2}- Ns(s+1)]. \label{1}
\end{equation}
If the interaction strength $A$ is a random variable with zero mean, then the
ground state of the system will have equal, $f_{0}=f_{S_{max}}=1/2$, probabilities to
have spin $S=0$ or $S=S_{max}$ (antiferromagnetism or
ferromagnetism). A similar situation takes place in the degenerate pairing
model \cite{racah} where the pair creation, $P_{0}^{\dagger}$, pair
annihilation, $P_{0}$, and particle number, $N$, operators form an SU(2)
pseudospin algebra. Then the eigenenergy is simply proportional to the pairing
constant so that, for a random sign of this constant, the ground state
pseudospin will be 0 (unpaired state of maximum seniority) or maximum possible 
(fully paired state of zero seniority), on average in 50\% of cases. In the Elliott SU(3) model \cite{ell}, as well as in any model with a rotational spectrum, the normal or inverted bands will happen evenly if the moment of
inertia takes positive or negative values randomly.

Let us consider a system of interacting fermions. For simplicity we limit
ourselves here to a case of $N$ identical particles on a single-$j$ 
shell which provides a generic framework for the extreme limit of strong
residual interaction. Rotational invariance is preserved,
so that all single-particle $m$-states are 
degenerate in energy. Within this space, the general
two-fermion rotationally invariant interaction can be written as
\begin{equation}
H=\sum_{L\Lambda}V_{L}P^{\dagger}_{L\Lambda}P_{L\Lambda},  \label{2.1}
\end{equation}
where the pair operators with pair spin $L$ and its projection $\Lambda$
are defined as
\begin{equation}
P^{\dagger}_{L\Lambda}=\frac{1}{\sqrt{2}}\sum_{mn}C^{L\Lambda}_{mn}
a^{\dagger}_{m}a^{\dagger}_{n}, \quad P_{L\Lambda}=\frac{1}{\sqrt{2}}
\sum_{mn}C^{L\Lambda}_{mn}a_{n}a_{m};                          \label{1a}
\end{equation}
 and $C$ are the Clebsch-Gordan coefficients.             
Because of Fermi statistics, only even  $L$ values are allowed in the single-$j$ space. This fact was ignored in the attempt
\cite{b1} to construct the quasiparticle ensemble with identical distributions of the parameters $V_{L}$ in the particle-particle channel and the parameters $\tilde{V}_{K}$ for the same interaction
transformed to the particle-hole channel,
$H\sim\sum_{K\kappa}\tilde{V}_{K}(a^{\dagger}a)_{K\kappa}
(a^{\dagger}a)_{K\tilde{\kappa}}$ (the difference between the
interactions in the two channels was discussed long ago by Belyaev \cite{bel}, 
and served as a justification for an interpolating model ``pairing plus
multipole-multipole forces''). Since $K$ can
take both even and odd values, the number of parameters is different in
the two representations, and $\tilde{V}_{K}$ cannot be 
independent if $V_{L}$ are. 

\begin{figure}
\begin{center}
\epsfxsize=6.8in \epsfbox{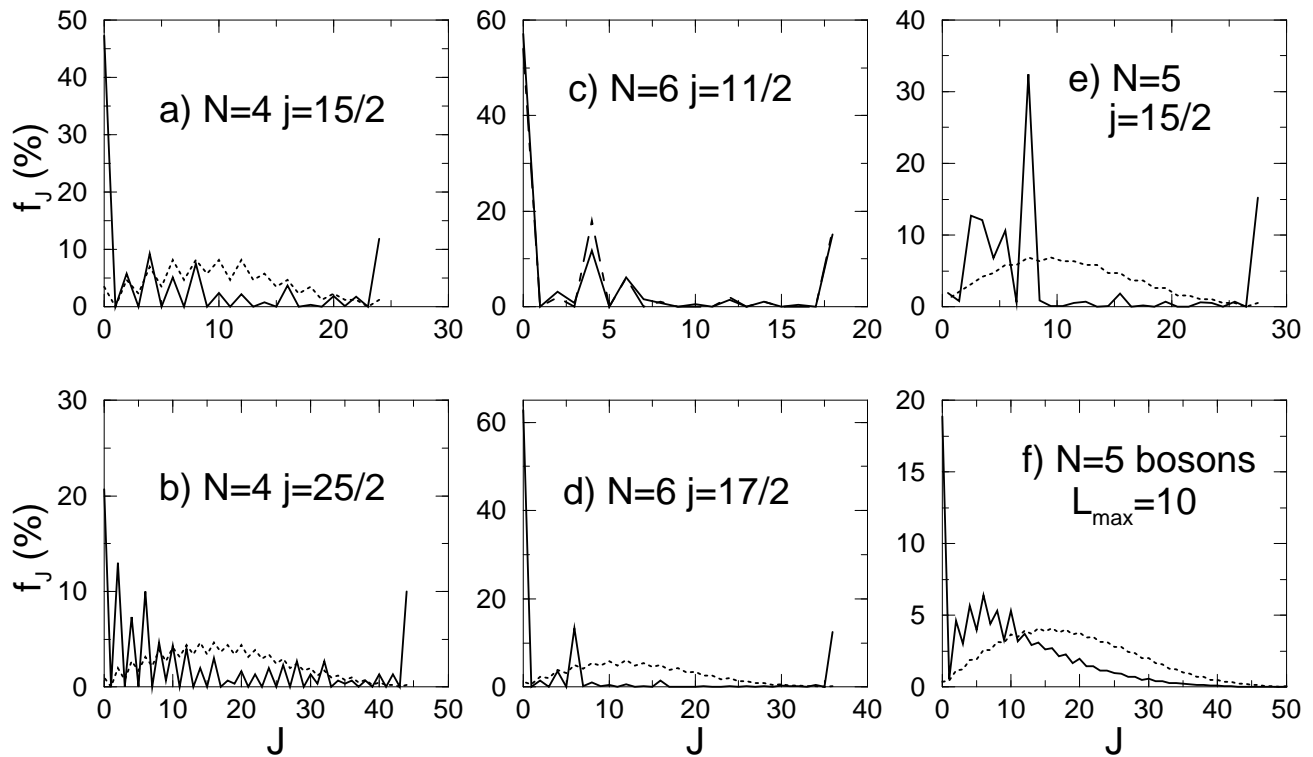}
\end{center}
\caption{
\label{fig1} The distribution of ground state angular momenta for various systems of $N$ fermions of spin $j$, ({\sl a--e}).The  bosonic approximation, $f_{J}^{b}$ is in panel ({\sl f}). The dotted lines are the statistical distribution of allowed $J$ and the solid lines are the ensemble results. In (c) the dashed line is for $V_{0} = 0$, i.e. no pairing.}
\end{figure}

Assuming that the coupling constants $V_{L}$ are random, uncorrelated and uniformly distributed between -1 and 1, we get the
distribution $f_J$ of the ground state spin $J_{0}$ shown in Fig. \ref{fig1}({\sl a--e}) for $N=4$ and $N=6$ at different values of $j$. For comparison we
show by dotted lines the statistical distributions 
$f^{s}_J$ based on the fraction of states of given $J$ in the entire Hilbert
space for given $N$. The overwhelming probability $f_{0}$ shows the same phenomenon in the uniform ensemble as observed earlier in  Gaussian ensembles of $V_{L}$ \cite{b1,b2,stu}.
Further evidence of the dominance of $J_{0}=0$ configurations
is given by the example,
Fig. \ref{fig1} {\sl(e)} , for an odd number of particles, where excess of the ground state 
spin $J_{0}=j$ is evidently related to the ground spin $J_{0}=0$ in the neighboring
even system. 

First we note that the effect seems to exist already in a crude
approximation modeling fermionic pairs by bosons. 
The commutation relations for the fermion pair operators (\ref{1a}) are ($L$ and
$L'$ are even),
 \begin{equation}
[P_{L'\Lambda'},P^{\dagger}_{L\Lambda}]=\delta_{L'L}
\delta_{\Lambda'\Lambda}+2\sum_{mm'n}
C^{L'\Lambda'}_{m'n}C^{L\Lambda}_{nm}a^{\dagger}_{m}a_{m'}.        \label{5}
\end{equation}
The second term in (\ref{5}) is of the order $N/\Omega$ where $\Omega$ is the
capacity ($=2j+1$ in our case) of the fermionic orbitals. It is small for 
a small number of fermions; for a nearly filled shell its effect is also small because of the particle-hole symmetry of states. For intermediate  shell occupation this term is not small but can be approximately substituted by its mean value (the monopole part with spin $K=0$). Then, after a simple renormalization,
$P_{L\Lambda}$ become bosonic operators.
This is  the assumption used in the original boson expansion
techniques \cite{bz,mk} and later in the interacting boson models: fermionic
pairs $P_{L\Lambda}$ are substituted by bosons $b_{L\Lambda}$,
and the Hamiltonian (\ref{2.1}) becomes a
sum of random bosonic energies $\sum_{L\Lambda}\omega_{L}n_{L\Lambda}$. The
ground state in each realization corresponds to the condensation of the bosons 
into the single-boson states $|L\Lambda)$ with the lowest value of $\omega_{L}$. For a given $L$, the many-boson states with different $J$ allowed for the condensate are degenerate, but the value $L=0$ is singled out by the obvious fact that for $\omega_{0}=\,$min
all degenerate states have total spin $J=0$ while for the minimum boson
energy $\omega_{L}$ at $L\neq 0$ any specific value of $J$, including $J=0$, 
appears only in a small fraction of states. If all $V_{L}$ have the same
distribution, we expect $f_{0}^{b}\approx 
1/k$ where $k$ is a number of (equiprobable) values of $L$. All other values
$J\neq 0$ appear with small probabilities $\sim 1/k^{2}$. This is
demonstrated by Fig. \ref{fig1}({\sl f}) where the pattern is qualitatively similar to that in Fig.   \ref{fig1}({\sl a--e}). The bosonic effect gives only a
part (decreasing with increasing $j$) of the $J_{0}=0$ dominance observed for the 
fermions. Another argument against the dominance of the bosonic correlations is
given in Fig. \ref{fig1}({\sl c}). Here we see that after exact elimination of the monopole term ($V_{L=0}\equiv 0$), the picture does not significantly
change although the value $V_{0}$ is now the lowest only in a small fraction, 
$\sim 2^{-(k-1)}$, of all cases (when all $V_{L\neq 0}$ are positive). 

%\begin{figure}
%\begin{center}
%\epsfxsize=8.0cm \epsfbox{ib_5.eps}
%\end{center}
%\caption{
%\label{fig2} $f_{J}^{b}$ for the bosonic approximation. Inset shows area about the origin; for 6 (11) possible values of $L$, $f_{0}>$ 1/6 (1/11). }
%\end{figure}

In our opinion, the main effect comes from the statistical correlations of the 
fermions.
They resolve the bosonic degeneracy in favor of the $J=0$ and $J=J_{max}$ ground
states. In the strong mixing among nearly degenerate states, the eigenstates
emerge as complicated chaotic superpositions. The only constraints left are the
conservation laws for the particle number and total spin. The latter can be
taken into account by the standard cranking approach \cite{erik,BM,gupta}. 
Thus, we model the system
by the Fermi-gas in statistical equilibrium with the occupation numbers $n_{m}$
of individual orbitals characterized by the angular momentum projection $m$
onto the cranking axis. The presence of the constraints creates a ``body-fixed
frame" and splits effective quasiparticle energies, although instead of the
collective rotation around a perpendicular axis we have here a random coupling
of individual spins with the symmetry (cranking) axis being the only direction which is
singled out in the system \cite{good}. Under the constraints 
\begin{equation}
        N=\sum_{m}n_{m}, \quad M=\sum_{m}mn_{m},                  \label{p6}
\end{equation}
 equilibrium statistical mechanics leads to the Fermi-Dirac
distribution
\begin{equation}
n_{m}=\frac{1}{\exp(\gamma m-\mu)+1}                        \label{p7}
\end{equation}
determined by the Lagrange multipliers of the chemical potential $\mu$
and cranking frequency $\gamma$; in the end the total projection $M$
(equivalent to the $K$ quantum number for axially deformed nuclei) is 
identified with the total spin $J$. 

The quantities $\mu(N,M)$ and $\gamma(N,M)$ can be found directly from
(\ref{p6}). At $M=0$ we have $\gamma=0$, so that the expansion in powers of
$\gamma$ allows one to study the most important region around $M=0$; the power expansion is sufficient for all $M$ except for the 
edges. With no cranking, one has the uniform distribution of occupancies $n^{0}_{m}=\bar{n}=N/\Omega$. With the 
perturbational cranking, the occupation numbers are
\begin{equation}
n_{m}=\bar{n}\left[1-\gamma m(1-\bar{n})+\frac{\gamma^{2}}{2}
(m^{2}-\langle m^{2}\rangle)(1-\bar{n})
(1-2\bar{n})+\;\cdots\right].                                       \label{p8}
\end{equation}
Here $\langle m^{2}\rangle=(1/\Omega)\sum_{m}m^{2}={\bf j}^{2}/3$, and terms of
higher orders are not shown explicitly.
The expectation value of energy in our statistical system can be written as
\begin{equation}
\langle H\rangle=\sum_{L\Lambda m_{1}m_{2}}V_{L}|C^{L\Lambda}_{m_{1}m_{2}}|^{2}
\langle n_{m_{1}}n_{m_{2}}\rangle.                                  \label{p9}
\end{equation}
Neglecting the correlations between the occupation numbers, $\langle n_{m_{1}}
n_{m_{2}}\rangle\approx n_{m_{1}}n_{m_{2}}$, we use the statistical result
(\ref{p8}) and calculate the geometrical sums with the Clebsch-Gordan
coefficients. Expressing the parameter $\gamma$ in terms of the total spin
$M\rightarrow J$, we come to the result including the terms of the second
order in $J^{2}$,
\begin{equation}
\langle H\rangle_{N,J}=\sum_{L}(2L+1)V_{L}[h_{0}(L)+h_{2}(L)J^{2}+h_{4}
(L)J^{4}],                                                  \label{p10}
\end{equation}
where 
\begin{equation}
h_{0}(L)=\bar{n}^{2}, \quad h_{2}(L)=\frac{3}{2}\,\frac{{\bf L}^{2}-2
{\bf j}^{2}}{{\bf j}^{4}\Omega^{2}},                               \label{p11}
\end{equation}
\begin{equation}
h_{4}(L)=\frac{9}{40}\,\frac{(1-2\bar{n})^{2}(3{\bf L}^{4}+3{\bf L}^{2}-
12{\bf j}^{2}{\bf L}^{2}-6{\bf j}^{2}+8{\bf j}^{4})}{(1-\bar{n})^{2}
N^{2}\Omega^{2}{\bf j}^{8}}.                                      \label{p12}
\end{equation}

$J_{0}$ is determined by the ensemble distributions of
$h_{2,4}=\sum_{L}(2L+1)V_{L}h_{2,4}(L).$                         
For all realizations of the random interaction with non-negative $h_{2}$ and
positive $h_{4}$, the ground state has spin $J_{0}=0$. If $h_{2}>0$ but $h_{4}
<0$, one has a local minimum of energy at $J=0$ although there is a possibility
to reach the absolute energy minimum at $J_{max}=(1/2)N
(\Omega-N)$. This will not happen if at $J=J_{max}$ we still have $h_{2}+
J_{max}^{2}h_{4}>0$. Therefore the probability to have the ground spin state equal
to zero turns out to be, in this approximation,
\begin{equation}
f_{0}=\int_{S(h_{2},h_{4})} dh_{2}dh_{4}{\cal P}(h_{2}h_{4}),     \label{p13}
\end{equation}
where the region $S$ is defined by the conditions $h_{2}>0, h_{4}>-(h_{2}/
J_{max}^{2})$. Since $h_{4}$ is small, the result is close to that for the right semi-plane $h_{2}>0$, and 
$f_{0}$ should be close to 50\%. For a Gaussian distribution of the parameters $V_{L}$ with zero mean and
variances $\sigma_{L}$, the distribution of the linear combinations $h_{2,4}$ is again Gaussian, and the integral over the region $S$ in (\ref{p13}) gives for this case
\begin{equation}
f_{0}=\frac{1}{4}+\frac{1}{2\pi}{\rm arctan}
\left[\frac{D+A/J_{max}^{2}}{\sqrt{AB-D^{2}}}\right],         \label{p16}
\end{equation}
which is close to 1/2. 
Here we introduced the combinations of geometric factors weighted with the
corresponding variances,
\begin{equation}
A=\sum_{L}(h_{2}(L))^{2}\sigma_{L}^{2}, \quad D=\sum_{L}h_{2}(L)h_{4}(L)
\sigma_{L}^{2}, \quad B=\sum_{L}(h_{4}(L))^{2}\sigma_{L}^{2}.  \label{p15}
\end{equation}

\begin{figure}
\begin{center}
\epsfxsize=8.0cm \epsfbox{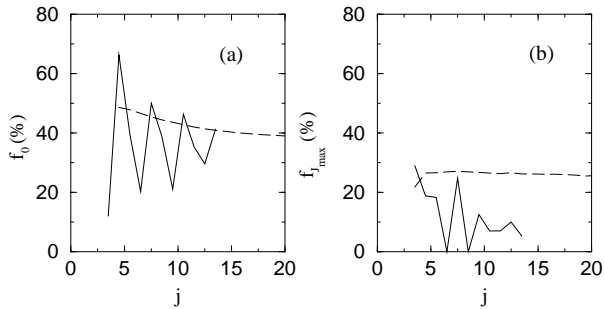}
\end{center}
\caption{
\label{exactstat}$f_{0}$ for $N=4$ and different $j$; ensemble results (solid line), statistical theory (dotted line, panel ({\sl a}); upper limit for $f_{J_{max}}$ from the  statistical theory and analysis of the $J_{max}$ region (dotted line, panel ({\sl b})).}
\end{figure}

The $\gamma$-expansion fails for large momenta. However, the states with high $M$ can be constructed exactly.
For Fig.  \ref{exactstat}  we used our statistical approach near $J=0$ in conjunction with the exact values in the end region $J=J_{max}$ to improve the above result for $f_{0}$ and to get an upper bound for $f_{J_{max}}$. Thus the statistical approach provides a good estimate for the dominance of $J=0$ and $J=J_{max}$ in the ground state; more subtle effects such as odd-even staggering should be considered separately.

\begin{figure}
\begin{center}
\epsfxsize=8.0cm \epsfbox{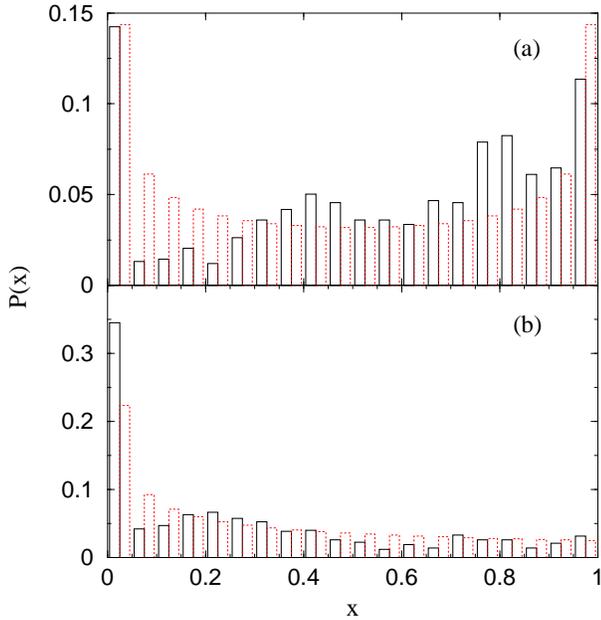}
\end{center}
\caption{
\label{fig4} The distribution 
of overlaps of $J_{0}=0$ ground states of the degenerate pairing model ($V_{0}=-1,\, V_{L\neq 0}=0$)  with those  for the ensemble choices a) random ensemble with $V_{0}=-1$, b) random ensemble. $N=6$ and $j=11/2$ in both cases, and the dotted line is the predicted $P(x)$.}
\end{figure}

Although the energy spectra with random two-body interactions bear clear
resemble the ordered spectra of pairing forces, the structure of
the eigenstates is close to that expected for 
chaotic dynamics \cite{baps2}. Fig. \ref{fig4}(b) shows the distribution $P(x)$
of the overlaps $x=|\langle J=0,{\rm g.s.}|0,{\rm p}\rangle|^{2}$ of ground
states with spin 0 obtained in the random ensemble with the ground state 
$|0,{\rm p}\rangle$ for the degenerate pairing model,
the latter corresponding to the
case of fixed $V_{0}=-1,\, V_{L\neq 0}=0$. In the chaotic limit the wave
functions are expected \cite{brody,BM} to behave as random superpositions of
basis states with uncorrelated components $C$ uniformly spread over a unit 
sphere, $P(C)\propto\delta(\sum C^{2}-1)$. This is equivalent to the 
distribution of a single component $P(C_{1})\propto (1-C_{1}^{2})^{(n-3)/2}$
where $n$ is the space dimension. For $n\gg 1$, the distribution $P(C_{1})$
is close to Gaussian whereas the overlaps $x=C_{1}^{2}$ obey the
Porter-Thomas distribution.
In the case of Fig. \ref{fig4} ($N=6$ particles, $j= 11/2$) the dimension of the  $J=0$ space is small, $n=3$, so that $P(C_{1})$ is constant, and we expect $P(x) \propto 1/\sqrt{x}$, as in the case of the pion multiplicity for the disordered chiral condensate. Another case considered in Fig.  \ref{fig4}(a) corresponds to the overlap of the degenerate pairing model ground state with the ground state in the model with $V_{0}=-1$,$V_{L\neq 0}$ random. Of course, here the completely
paired state can appear as the ground state even for random strengths in the
channels $L\neq 0$ which gives the peak at the overlap $x=1$. But the 
character of the distribution changes as well becoming effectively
two-dimensional: for $n=2$, $P(x)\propto 1/\sqrt{x(1-x)}$. 

To conclude, we have shown that statistical correlations of fermions in a 
finite Fermi system with random interactions 
drive the ground state spin to its minimum or maximum
value. This effect is related to the geometrical chaoticity of the random spin coupling of individual particles. This means, that the dominance of $0^{+}$ ground states in even-even nuclei may at least partly come from incoherent interactions rather than solely from coherent pairing.
The structure of ground states with an ``antiferromagnetic" type ordering, 
$J_{0}=0$, is compatible with the predictions for chaotic dynamics.
Quantitative relations between the effects of geometric chaoticity and pure
dynamic effects in finite many-body systems should be an interesting subject for further detailed studies. \\
\\
The authors wish to acknowledge  P. Cejnar whose expertise in the interacting boson model was very helpful. The authors are grateful to  G.F. Bertsch, B.A. Brown, V. Cerovski, V.V. Flambaum, M. Horoi,  
F.M. Izrailev, and D. Kusnezov for constructive discussions.
This work was supported by the NSF grant 96-05207.

\end{document}